\documentclass[aoas,nameyear,seceqn,mtplusscr,dvips]{arximspdf}
\usepackage{euscript}
\usepackage{graphicx,mathrsfs}

\doi{10.1214/09-AOAS245}
\volume{4}
\issue{2}
\pubyear{2010}
\firstpage{1034}
\lastpage{1055}

\makeatletter
\newtheorem{theorem}{Theorem}
\newtheorem{corollary}{Corollary}
\newtheorem{proposition}{Proposition}
\newproclaim{remark}{Remark}
\newproclaim{example}{Example}
\makeatother

\begin{document}
\begin{frontmatter}

\title{DISCO analysis: A nonparametric extension of analysis of variance}
\runtitle{Disco analysis}

\begin{aug}
\author[a]{\fnms{Maria L.} \snm{Rizzo}\ead[label=e1]{mrizzo@bgsu.edu}\corref{}}
\and
\author[a]{\fnms{G\'{a}bor J.} \snm{Sz\'{e}kely}\thanksref{t2}\ead[label=e2]{gabors@bgsu.edu}}
\runauthor{M. L. Rizzo and G. J. Sz\'{e}kely}
\thankstext{t2}{Research supported by the National Science Foundation,
while working at the Foundation:
Program Director, Statistics \& Probability, 2006--2009.}
\affiliation{Bowling Green State University}
\address[a]{Department of Mathematics and Statistics\\
Bowling Green State University\\
Bowling Green, Ohio 43403\\
USA\\
\printead{e1}\\
\phantom{E-mail: }\printead*{e2}}
\end{aug}

\received{\smonth{5} \syear{2008}}
\revised{\smonth{3} \syear{2009}}

%
\begin{abstract}
In classical analysis of variance, dispersion is measured by
considering squared distances of sample elements from the sample
mean. We consider a measure of dispersion for univariate or
multivariate response based on all pairwise distances between-sample
elements, and derive an analogous distance components (DISCO)
decomposition for powers of distance in $(0,2]$. The ANOVA F statistic
is obtained when the index (exponent) is 2. For each index in $(0,2)$,
this decomposition determines a nonparametric test for the
multi-sample hypothesis of equal distributions that is
statistically consistent against general alternatives.
\end{abstract}

%

\end{frontmatter}

\section{Introduction}

In classical analysis of variance (ANOVA) and multivariate analysis
of variance (MANOVA), the $K$-sample hypothesis for equal means is
%
\begin{equation}
H_0\dvtx  \mu_1=\cdots=\mu_K \label{Hanova}
\end{equation}
vs $H_1\dvtx \mu_j \neq\mu_k, $ for some $j \neq k$, where
$\mu_1,\dots,\mu_K$ are the means or mean vectors of the $K$ sampled
populations. Inference requires that random error is normally
distributed with mean zero and constant variance (see, e.g.,
\cite{cc57}, \cite{scheffe53}, \cite{scn92}, \cite{ht87}, or
\cite{mkb79}).

Analysis of variance partitions the total variance of the observed
response variable into $\mathit{SST}$ (sum of squared error due to
treatments) and $\mathit{SSE}$ (sum of within-sample squared error). When the
usual assumptions of normality and common error variance hold, under
the null hypothesis distributions are identical, and under the
alternative hypothesis distributions differ only in location (are
identical after translation). If distributions differ in
\textit{location only}, for a univariate response, methods based on
ranks such as the nonparametric Kruskal--Wallis test or Mood's median
test can be applied to test the hypothesis of equal population
medians (see, e.g., \citeauthor{hw99} (\citeyear{hw99}, Chapter~6)).

In case the assumptions of normality or common variance do not\break hold,
one could apply $F$ statistics via a permutation test procedure\break
(\citeauthor{et93} (\citeyear{et93}, Chapter~15), \citeauthor{dh97} (\citeyear{dh97}, Chapter~4)). However, in practice,
distributions with equal means may differ in other characteristics,
while $F$ statistics test the hypothesis (\ref{Hanova}) of equal means.

We extend ANOVA and MANOVA to testing the more general hypothesis (\ref
{Hk}) with the help of a decomposition for other exponents than squared
distance.

For $K$ independent random samples from distributions with
cumulative distribution function (c.d.f.) $F_1,\dots,F_K$ respectively,
the $K$-sample hypothesis for equal distributions is
%
\begin{equation}
H_0\dvtx  F_1=\cdots=F_K \label{Hk}
\end{equation}
versus the composite alternative $F_j \neq F_k$ for some $1 \leq j <
k \leq K$. Here each of the $K$ random variables are assumed to take
values in $\mathbb R^p$ for some integer $p\geq1$, and the
distributions $F_j$ are unspecified.

We propose a new method, called \textit{distance components} (DISCO), of
measuring the total dispersion of the samples, which admits a
partition of the total dispersion into components analogous to the
variance components in ANOVA. The resulting distance components
determine a test for the more general hypothesis (\ref{Hk}) of equal
distributions. We introduce a measure of dispersion based on
Euclidean distances between all pairs of sample elements, for any
power $\alpha$ of distances such that $\alpha\in(0,2]$, hereafter
called the \textit{index}. The usual ANOVA decomposition of the total
squared error is obtained as the special case $\alpha=2$. For all
other values of the index $0 < \alpha< 2$, we obtain a
decomposition such that the corresponding ``F'' statistic determines
a test of the general hypothesis (\ref{Hk}) that is statistically
consistent against general alternatives.

\cite{aa94} proposed a general model for structured data where the
distribution of the response variable is modeled in terms of
distributions. A~hypothesis of no treatment effect or no interaction
effect is that the corresponding distribution term in the model is
identically zero. For an overview, see \cite{bp01} and the references therein.

Other distance based approaches to testing (\ref{Hanova}) or (\ref
{Hk}) have been proposed in recent literature by \cite{gk99} and \cite
{anderson01}, with applications in ecology, economics, and genetics
(\cite{ma01}; \cite{esq92}; \cite{zs06}). These methods differ from our proposed
approach in that they employ the squared distance (and thus test a
different hypothesis), a different way of decomposing the distances, or
a dissimilarity measure other than powers of Euclidean distances.

Our main results, the statistics for measuring distances between
samples and the method of partitioning the total dispersion, are
introduced in Section \ref{sec2}. Properties of these statistics and the
proposed DISCO test for the general hypothesis (\ref{Hk}) are
presented in Section \ref{sec3}, and DISCO decomposition for multi-factor
models follows in Section \ref{s:DD}.
Implementation, examples, and empirical results are covered
in Sections \ref{s:ie} and \ref{empirical}.

\section{Distance components}\label{sec2}

\subsection{DISCO statistics}

Define the empirical distance between distributions as follows. For
two samples $A= \{ a_1,\dots,a_{n_1} \}$ and $B=\{ b_1,\dots,b_{n_2}
\}$, the $d_\alpha$-distance between $A$ and $B$ is defined as
\[
d_\alpha(A,B) = \frac{n_1 n_2}{n_1+n_2}
[2g_\alpha(A,B)-g_\alpha(A,A)-g_\alpha(B,B)],
\]
where
%
\begin{equation}\label{g}
g_\alpha(A,B) = \frac{1}{n_1 n_2}
\sum\limits_{i=1}^{n_1} \sum\limits_{m=1}^{n_2} \|a_i-b_m\|^\alpha
\end{equation}
is a version of the Gini mean distance statistic and $\| \cdot\|$
denotes the Euclidean norm. The constant $\frac{n_1n_2}{n_1+n_2}$ is
half the harmonic mean of the sample sizes.

In the special case $\alpha=2$, the $d_2$-distance for a univariate
response variable measures variance, and there is an interesting
relation between the $d_2$-distances and the ANOVA sum of squares
for treatments. The details are explored below.

\begin{proposition} \label{P:sst2}
Let $A=\{a_1,\dots,a_{n_1}\}$ and $B=\{b_1,\dots,b_{n_2}\}$ with
means $\bar a$ and $\bar b$ respectively. Then
\[
d_2(A,B)=2  \mathit{SST} = 2[n_1(\bar a - \bar c)^2 +
n_2 (\bar b - \bar c)^2],
\]
where $\bar c = (n_1 \bar a + n_2 \bar b) / (n_1 + n_2)$.
\end{proposition}

The proof of Proposition \ref{P:sst2} is given in the \hyperref[append]{Appendix}.

In the following, $A_1,\dots,A_K$ are $p$-dimensional samples with
sizes $n_1,\dots,$ $n_K$ respectively, and $N=n_1+\cdots+ n_K$.

The $K$-sample $d_\alpha$-distance statistic that takes the role of
ANOVA sum of squares for treatments is the weighted sum of
dispersion statistics. For the balanced design with common sample
size $n$, define the between-sample dispersion as
%
\begin{equation}\label{between}
S_\alpha= S_\alpha(A_1,A_2,\dots,A_K) =
\frac1K \sum\limits_{1 \leq j < k \leq K}  d_\alpha(A_j,A_k).
\end{equation}

For unbalanced designs with sample sizes $n_1,n_2,\dots,n_K$, for
each pair of samples the factor $1/K=n/N$ in (\ref{between}) is
replaced by $\tilde n_{jk}/N$, where $\tilde n_{jk}$ is the
arithmetic mean of $n_j$ and $n_k$. Thus, for the general case the
between-sample dispersion is
\begin{eqnarray}\label{betweenjk}
S_\alpha&=& S_\alpha(A_1,A_2,\dots,A_K) =
\sum\limits_{1 \leq j < k \leq K}
\biggl(\frac{n_j + n_k}{2N}\biggr)
 d_\alpha(A_j,A_k) \nonumber\\[-8pt]\\[-8pt]
 &=&
\sum\limits_{1 \leq j < k \leq K} \biggl\{\frac{n_j n_k}{2N} \bigl(
2g_\alpha(A_j,A_k)-g_\alpha(A_j,A_j)-g_\alpha(A_k,A_k ) \bigr)
\biggr\}. \nonumber
\end{eqnarray}
Note that if $K=2$, $p=1$, and $\alpha=2$, we have
$S_2=d_2(A_1,A_2)/2 = \mathit{SST}.$

It follows from Theorem 1 in the following section that for all
$0<\alpha<2$ the statistic $S_\alpha$ determines a statistically
consistent test for equality of distributions.

First let us explore the relation between $S_2$ and $\mathit{SST}$. A well-known
$U$-statistic is the sample variance $S^2$. If $x_1,\dots,x_n$
is a sample, then
%
\begin{equation}
\frac{1}{{n \choose{2}}}\sum\limits_{1\leq i < m \leq n} \frac12
(x_i-x_m)^2 = S^2 = \frac{1}{n-1} \sum\limits_{i=1}^n (x_i-\bar x)^2.
\label{var}
\end{equation}
This example is given by \cite{serfling80}, page 173. Notice that
if $A_1,\dots,A_K$ have common sample size $n$, then (\ref{id:d2})
and (\ref{var}) can be applied to compute
\begin{eqnarray*}
S_2 &=& \frac{1}{K} \sum\limits_{1 \leq j < k \leq K}
d_2(A_j,A_k) =
\frac{K-1}{{K \choose{2}}} \sum\limits_{1 \leq j < k \leq K}
\frac12
n (\bar a_{\cdot j} - \bar a_{\cdot k})^2 \\
&=& \sum\limits_{j=1}^K n(\bar
a_{\cdot j}-\bar a_{\cdot \cdot })^2
= \mathit{SST}.
\end{eqnarray*}
In the case of arbitrary sample sizes, the same relation holds: $S_2
= \mathit{SST}$. This identity is obtained as a corollary from the
decomposition of total dispersion into the between and within
components, which follows in Section \ref{s:decomp}.

\subsection{DISCO decomposition\label{s:decomp}}

Define the total dispersion of the observed response by
%
\begin{equation}\label{total}
T_\alpha= T_\alpha(A_1,\dots,A_K) = \frac{N}{2}  g_\alpha(A,A),
\end{equation}
where $A = \sum_{j=1}^K A_j$ is the pooled sample and $g_\alpha$ is
given by (\ref{g}). Similarly, define the within-sample dispersion
statistic
%
\begin{equation}\label{within}
W_\alpha= W_\alpha(A_1,\dots,A_K) = \sum\limits_{j=1}^K
\frac{n_j}{2}  g_\alpha(A_j, A_j).
\end{equation}
Then if $0 < \alpha\leq2$, we have the decomposition $T_\alpha=
S_\alpha+ W_\alpha$, where both $S_\alpha$ and $W_\alpha$ are nonnegative.
Moreover, for $0<\alpha<2$, $S_\alpha= 0$ if and only if $A_1= \cdots
=A_K$. For the proof, we need the following definition and theorem.

Suppose that $X$ and $X'$ are independent and identically
distributed (i.i.d.), and $Y$ and $Y'$ are i.i.d., independent of $X$. If
$\alpha$ is a constant such that $E\|X\|^\alpha< \infty$ and
$E\|Y\|^\alpha< \infty$, define the $\EuScript{E}_\alpha$-distance
(\textit{energy} distance) between the distributions of $X$ and $Y$
as
\[
\EuScript{E}_\alpha(X,Y) =
2E\|X-Y\|^\alpha- E\|X-X'\|^\alpha- E\|Y-Y'\|^\alpha.
\]

\begin{theorem} \label{Th1}
Suppose that ${X}$ and ${X}' \in\mathbb R^p$ are i.i.d. with
distribution $F$, ${Y}$ and ${Y}' \in\mathbb R^p$ are i.i.d. with
distribution $G$, and $Y$ is independent of $X$. If $0<\alpha\leq
2$ is a constant such that $E\|{X}\|^\alpha<\infty$ and
$E\|{Y}\|^\alpha<\infty$, then the following statements hold:
\begin{longlist}
\item[(i)] $\EuScript{E}_\alpha(X,Y)
\geq0$.

\item[(ii)] If $0<\alpha<2$, then $\EuScript{E}_\alpha(X,Y) = 0$
if and only if $X \stackrel{\mathscr D}= Y$.

\item[(iii)] If $\alpha= 2$, then ${\mathscr E_\alpha}(X,Y) = 0$
if and only if $E[X]=E[Y]$.
\end{longlist}
\end{theorem}

\begin{pf}
The proof for multivariate samples is given in \citeauthor{sr05b} (\citeyear{sr05b}). Here
we present a more elementary proof for the univariate case.

First consider the case $0 < \alpha< 2$. Using the fact that
$|X-Y|^\alpha$ is a nonnegative random variable, and making the
substitution $u=t^{1/\alpha}$, we have
\begin{eqnarray*}
E|X-Y|^\alpha&=& \int_0^\infty P(|X-Y|^\alpha> t)\,dt =
\int_0^\infty P(|X-Y| > t^{1/\alpha}) \,dt \\
&=&
\int_0^\infty\alpha u^{\alpha-1} P(|X-Y| > u) \, du \\
&=&
\int_{\mathbb R} \alpha|u|^{\alpha-1}[P(X < u < Y) + P(Y < u <
X)]\, du \\
&=&
\int_{\mathbb R} \alpha|u|^{\alpha-1}\bigl[F(u)\bigl(1-G(u)\bigr) +
G(u)\bigl(1-F(u)\bigr)\bigr] du.
\end{eqnarray*}
Similarly,
\begin{eqnarray*}
E|X-X'|^\alpha&=&
\int_{\mathbb R} 2\alpha|u|^{\alpha-1}\bigl[F(u)\bigl(1-F(u)\bigr)\bigr] \,du, \\
E|Y-Y'|^\alpha&=&
\int_{\mathbb R} 2\alpha|u|^{\alpha-1}\bigl[G(u)\bigl(1-G(u)\bigr)\bigr] \,du.
\end{eqnarray*}
Thus,
%
\begin{eqnarray} \label{FG2}
&&2\alpha\int_{\mathbb R}  |u|^{\alpha-1}\bigl(F(u)- G(u)\bigr)^2\,du \\
&&\qquad =
2\alpha\int_{\mathbb R} |u|^{\alpha-1} \bigl[F(u)\bigl(1-G(u)\bigr) +
\bigl(1-F(u)\bigr)G(u)\nonumber
\\
&&\hspace*{90pt}{}-F(u)\bigl(1-F(u)\bigr) -G(u)\bigl(1-G(u)\bigr)\bigr]\,du \nonumber\\
&&\qquad = 2E|X-Y|^\alpha- E|X-X'|^\alpha- E|Y-Y'|^\alpha= \EuScript{E}_\alpha(X,Y). \label{FG2R}
\end{eqnarray}
The integral (\ref{FG2}) converges to a non-negative constant if
$|\alpha-1|<1$. Hence, (\ref{FG2R}) is non-negative and finite for all
$0<\alpha<2$. A necessary and sufficient condition that
(\ref{FG2R}) equals zero is that $F=G$ a.e., and $X \stackrel
{\mathscr D}= Y$. This proves (i) and (ii) for the case
$0<\alpha<2$.

Finally, for the case $\alpha=2$, we have
\begin{eqnarray*}
\EuScript{E}_2(X,Y) &= &2E|X-Y|^2-E|X-X'|^2-E|Y-Y'|^2 \\
& =& 2(E[X] -
E[Y])^2 \geq0,
\end{eqnarray*}
with equality if and only if $E[X] = E[Y]$.
\end{pf}

A consequence of Theorem \ref{Th1} is that the empirical distance
between samples is always non-negative:

\begin{corollary}\label{c1}For all $p$-dimensional samples $A_1,\dots,A_K$,
$K \geq2$, and \mbox{$0 < \alpha\leq2$}, the following statements hold:
\begin{longlist}
\item[(i)] $S_\alpha(A_1,\dots,A_K) \geq0$.
\item[(ii)] If $0 < \alpha< 2$, then $S_\alpha(A_1,\dots,A_K)=0$ if
and only if
$A_1=\cdots=A_K$.
\item[(iii)] $S_2(A_1,\dots,A_K)=0$ if and only if $A_1,\dots,A_K$
have equal means.
\end{longlist}
\end{corollary}

\begin{pf}
Let $A_j=\{a_1,\dots,a_{n_j}\}$ and $A_k=\{b_1,\dots,b_{n_k}\}$.
Define i.i.d. random variables $X$ and $X'$ uniformly distributed on
$A_j$, and define i.i.d. random variables $Y$ and $Y'$ uniformly
distributed on $A_k$. Then $E\|X-Y\|^\alpha= g_\alpha(A_j,A_k)$,
$E\|X-X'\|^\alpha=g_\alpha(A_j,A_j)$, $E\|Y-Y'\|^\alpha=
g_\alpha(A_k,A_k),$ and
\[
\frac{n_1n_2}{n_1+n_2} \EuScript{E}_\alpha(X,Y) = d_\alpha(A_j,A_k).
\]
Hence, for all $0 < \alpha\leq2$, Theorem \ref{Th1}(i) implies that
$S_\alpha(A_j,A_k) \geq0$. If $0 < \alpha< 2$, then by Theorem
\ref{Th1}(ii) equality to zero holds if and only if $X \stackrel
{\mathscr D}= Y$ (if and only if $A_j=A_k$). This proves (i) and
(ii) for the case $K=2$, and the result for $K \geq2$ follows by
induction. Statement (iii) follows from Theorem \ref{Th1}(iii).
\end{pf}

Our next theorem is the DISCO decomposition of total dispersion into
between-sample and within-sample components.

\begin{theorem} \label{decomp}
For all integers $K \geq2$, the total dispersion $T_\alpha$
(\ref{total}) of $K$ samples can be decomposed as
\[
T_\alpha(A_1,\dots,A_K)=S_\alpha(A_1,\dots,A_K) +
W_\alpha(A_1,\dots,A_K),
\]
where $S_\alpha\geq0$ and $W_\alpha\geq0$ are the between-sample
and within-sample measures of dispersion given by (\ref{betweenjk})
and (\ref{within}), respectively.
\end{theorem}

\begin{pf}
Let $G_{jk}=n_j n_k g_\alpha(A_j,A_k)$, and
$g_{jk}=g_\alpha(A_j,A_k)$. First consider the balanced design, with
common sample size $n$. In this case $(n_jn_k)/(n_j+n_k)=n/2$ and
$S_\alpha$ can be computed by (\ref{between}), so that
\begin{eqnarray*} \nonumber
T_\alpha- S_\alpha&=& \frac N2 g(A,A) - \frac1K \sum\limits_{j <
k} \frac n2 (2g_{jk}-g_{jj}-g_{kk})
\\&=& \frac{1}{2N} \biggl( \sum_j G_{jj}
+ \sum_{j<k} 2 G_{jk} \biggr) - \frac n{2K}
\sum_{j<k} \frac1{n^2}(2G_{jk}-G_{jj}-G_{kk}) \\&=&
\frac{1}{2N} \biggl( \sum_j G_{jj}
+ \sum_{j<k} (G_{jj}+G_{kk}) \biggr) \\&=&
\frac{1}{2N} \biggl( \sum_j G_{jj}
+ (K-1) \sum_{j} G_{jj} \biggr) \\&=&
\frac K{2N} \sum_j n^2 g_{jj}
= \frac{1}{2n} \sum n^2 g_{jj} = \frac12 \sum_j n g_{jj} = W_\alpha.
\end{eqnarray*}
The proof for the general case is similar; the details are given in
the \hyperref[append]{Appendix}.
\end{pf}

\begin{corollary}
If $p=1$, then for all integers $K \geq2$ the between-sample
dispersion $S_2$ for $K$ samples is equal to SST, and the $\alpha=2$
decomposition of total dispersion $T_2=S_2+W_2$ is exactly the ANOVA
decomposition of the total squared error: $\mathit{SS}(\mathit{total})=\mathit{SST}+\mathit{SSE}$.
\end{corollary}

\begin{pf}
Applying (\ref{var}) to the $\alpha=2$ Gini statistics shows that,
for samples $A_1, \dots, A_K$,
\[
g_2(A_j, A_j) = 2\hat\sigma_j^2,
\]
where $\hat\sigma_j^2 = n_j^{-1} \sum_{i=1}^{n_j}(a_{ij}-\bar
a_{.j})^2$, $j=1,\dots,K$. The within-sample sum of squares is
$\sum_{j=1}^K n_j \hat\sigma_j^2$. Similarly, the total sum of
squares is $N\hat\sigma^2=\frac{N}{2}g_2(A,A)$.

Thus, $W_2(A_1,\dots,A_K)=\sum_{j=1}^K n_j \hat\sigma^2_j=\mathit{SSE}$, and
$T_2(A_1,\dots,A_K)=N \hat\sigma^2 = \mathit{SS}(\mathit{total})$. Therefore, by the
ANOVA decomposition $\mathit{SS}(\mathit{total})=\mathit{SST}+\mathit{SSE}$ and Theorem \ref{decomp}, we
have
\[
S_2(A_1,\dots,A_K)=T_2(A_1,\dots,A_K) - W_2(A_1,\dots,A_K),
\]
hence, $S_2=\mathit{SST}$ and we obtain the one-way ANOVA decomposition of
total sum of squares.
\end{pf}

\section{DISCO hypothesis tests}\label{sec3}

Assume that $A_1,\dots,A_K$ are independent random samples of size
$n_1,\dots,n_K$ from the distributions of random variables
$X_1,\dots,X_K$ respectively.

\subsection{The DISCO $F_\alpha$ ratio for equal distributions}
\label{S:test}

Analogous to the ANOVA decomposition, under the null hypothesis of
equal distributions, $S_\alpha$ and $W_\alpha$ are both estimators
of the same parameter $E\|X_j-X_j'\|^\alpha$, where $X_j'$ and $X_j$
are i.i.d. The Gini mean $g_\alpha(A_j,A_j)$ is a biased estimator of
$E\|X_j-X_j'\|^{\alpha}$. An unbiased estimator of
$E\|X_j-X_j'\|^\alpha$ is $({n_j}/({n_j-1}))  g_\alpha(A_j,A_j)$.
Under the null hypothesis (\ref{Hk}) we have
\begin{eqnarray*}
E[S_\alpha]&=&
\frac{1}{2N} \sum_{1 \leq j<k \leq K} n_j n_k \biggl(2\xi- \frac
{n_j-1}{n_j} \xi-
\frac{n_k-1}{n_k} \xi\biggr)
\\&=& \frac{\xi}{2N} \sum_{1 \leq j<k \leq K}n_j n_k \biggl(
\frac{1}{n_j}+\frac{1}{n_k}\biggr)=\frac{K-1}{2} \xi,
\end{eqnarray*}
and
\[
E[W_\alpha] = \sum_{j=1}^K \frac{n_j}2 \biggl(\frac
{n_j-1}{n_j}\biggr) \xi
= \frac{N-K}2 \xi,
\]
where $\xi= E\|X_j-X_j'\|^\alpha$. Our proposed statistic for
testing equality of distributions is
\[
D_{n,\alpha} = F_\alpha=
\frac{S_\alpha/(K-1)}{W_\alpha/(N-K)}.
\]
Although in general $D_{n,\alpha}$ does not have an $F$
distribution, $F_\alpha$ has similar properties as the ANOVA $F$
statistic in the sense that $F_\alpha$ is non-negative and large
values of $F_\alpha$ support the alternative hypothesis. The details
of the decomposition can be summarized in a table similar to the
familiar ANOVA tables. See, for example, Tables \ref{t3} and \ref{tab1}.

\subsection{Permutation test implementation}\label{permtest}

The DISCO test can be implemented in a distribution free way by a
permutation test approach. Permutation tests are described in \citeauthor{et93} (\citeyear{et93}) and \citeauthor{dh97} (\citeyear{dh97}).
The~achieved significance level of a permutation test is exact.

Let $\nu=1$: \textit{N} be the vector of sample indices of the pooled sample
$A=(y_i)$, and let $\pi(\nu)$ denote a permutation of the elements
of $\nu$. The statistic $F_\alpha(A;\pi)$ is computed as
$F_\alpha(y_{\pi(i)})$. Under the null hypothesis (\ref{Hk}) the
statistics $F_\alpha(y_i)$ and $F_\alpha(y_{\pi(i)})$ are
identically distributed for every permutation $\pi$ of $\nu$.

\subsubsection*{Permutation test procedure}

\begin{enumerate}[iii.] \label{s:permproc}
\item[i.] Compute the observed test statistic $F_\alpha=F_\alpha(A;\nu)$.
\item[ii.] For each replicate, indexed $r=1,\dots,R$,
generate a random permutation $\pi_r=\pi(\nu)$
and compute the statistic $F^{(r)}_\alpha=F_\alpha(A;\pi_r)$.
\item[iii.] Compute the significance level (the empirical
$p$-value) by
\[
\hat p = \frac{1+\# \{F_\alpha^{(r)} \geq F_\alpha\} }{R+1}=
\frac{ \{ 1+\sum_{r=1}^R I(F_\alpha^{(r)} \geq F_\alpha)
\} }{R+1},
\]
where $I(\cdot)$ is the indicator function.
\end{enumerate}
The formula for $\hat p$ is given by \citeauthor{dh97} (\citeyear{dh97}, page~159), who
state that ``As a practical matter, it is rarely possible or necessary
to compute the permutation $P$-value exactly'' and ``at least 99 and at
most 999 random permutations should suffice.''

\subsection{Limit distribution}

For all $0<\alpha<2$, under the null hypothesis of equal
distributions, $d_\alpha(A_j,A_k)$ converges in distribution to a
quadratic form of centered Gaussian random variables (see details in
\citeauthor{sr05a} (\citeyear{sr05a}, \citeyear{sr05b})). Hence, under $H_0$ the mean between-sample
component $S_\alpha/(K-1)$ of the $F_\alpha$ ratio converges in
distribution to a quadratic form of centered Gaussian random
variables. The mean within-sample component $W_\alpha/(N-K)$
converges in probability to a constant by the law of large numbers.
Therefore, for all $0 < \alpha< 2$ by Slutsky's theorem under $H_0$,
the $F_\alpha$ ratio converges in distribution to a quadratic form
%
\begin{equation}\label{Q}
Q = \sum_{i=1}^\infty\lambda_i Z_i^2,
\end{equation}
where $Z_i$ are independent standard normal variables and
$\lambda_i$ are positive constants.

The DISCO test rejects (\ref{Hk}) if the test statistic $F_\alpha$
exceeds the upper percentile of the null distribution of $F_\alpha$
corresponding to the significance level $\alpha_0$. \cite{sb03}
proved that for quadratic forms (\ref{Q}) with $E[Q]=1$,
\[
P\bigl(Q \geq\bigl(\Phi^{-1}(1-\alpha_0/2)\bigr)^2\bigr) \leq\alpha_0,
\]
for $\alpha_0 \leq0.215$, where $\Phi(\cdot)$ is the standard
normal c.d.f.

\subsection{Consistency}

The advantage of applying an index in $(0,2)$ rather than squared
distances is that for exponents $0<\alpha<2$ all types of
differences between distributions are detected, and the test is
statistically consistent.

\begin{theorem}If $0<\alpha<2$, the DISCO test of the hypothesis (\ref{Hk})
is statistically consistent against all alternatives with finite
second moments.
\end{theorem}

\begin{pf}
Suppose that the null hypothesis is false. Then $F_j \neq F_k$ for
some $(j,k)$. Let $c>0$ be an arbitrary constant. We need to prove
that
\[
\lim\limits_{N \to\infty} P(F_\alpha> c) = 1.
\]
Here $N \to\infty$ is understood to mean that each $n_j \to
\infty$ and
\[
\lim_{n_1,\dots,n_K \to\infty} \frac{n_j}{n_1+\cdots+n_K} = p_j,
\qquad j=1,\dots,K,
\]
where $0<p_j<1$ and $\sum_{j=1}^K p_j = 1.$
Then
\begin{eqnarray*}
P(F_\alpha> c) & \geq& P\biggl(\frac{n_j+n_k}{2N}  \cdot
\frac{d_\alpha(A_j,A_k)}{K-1}  \cdot  \frac{N-K}{W_\alpha} >
c\biggr)
\\&=& P\biggl(d_\alpha(A_j,A_k) >
\frac{2cN(K-1)W_\alpha}{(n_j+n_k) (N-K)}\biggr).
\end{eqnarray*}
Statistical consistency of $d_\alpha(A_j,A_k)$ for $0<\alpha<2$
follows as a special case from \cite{sb03}. There are constants
$c_1$ and $c_2$ such that
\begin{eqnarray*}
\lim\limits_{N \to\infty} P(F_\alpha>c) &=&
\lim\limits_{N \to\infty} P\biggl(d_\alpha(A_j,A_k)>\frac
{2c(K-1)W_\alpha}{(p_1+p_2)(N-K)}\biggr)
\\&=&
\lim\limits_{N \to\infty} P\biggl(d_\alpha(A_j,A_k)>\frac
{c_1W_\alpha}{(N-K)}\biggr)
\\&=&
\lim\limits_{N \to\infty} P\bigl(d_\alpha(A_j,A_k)>c_2\bigr) =1
\end{eqnarray*}
by the statistical consistency of $d_\alpha(A_j,A_k)$.
\end{pf}

The corresponding $F_2$ statistic does not determine a consistent
test and does not necessarily detect differences of scale or other
characteristics.

\begin{remark}
A DISCO test is applicable even when first moments do not exist. For
any distribution such that an $\varepsilon$-moment exists, for some
$\varepsilon> 0$, we can choose $0<\alpha<\varepsilon/2$, which is
sufficient for statistical consistency because \mbox{$E\|X-Y\|^{2\alpha
}<\infty$}.
\end{remark}

\section{The DISCO decomposition in the general case\label{s:DD}}

Here we use the traditional formula notation from linear models. Let $Y
\sim A$ specify a completely randomized design on response $Y$ by
group variable (factor) $A$ with $a$ levels. If factor $B$ has $b$ levels,
and interaction $A\dvtx B$ denotes the crossed factors $A$ and $B$ with $ab$
levels, then $Y \sim A+B$ is the corresponding two-factor additive
model, and $Y \sim A * B = A+B+A\dvtx B$ is the two-way design with interaction.

Let $S(A)$, $W(A)$ denote the between and within components obtained by
a decomposition on factor $A$. In this section we omit the subscript
$\alpha$ when the expression is applicable for $0<\alpha\leq2$.

\subsection{The two-way DISCO decomposition}

Applying the theorem for DISCO decomposition to the model
$Y \sim A + B$, we have
\begin{eqnarray*}
T = S(A) + W(A) = S(B) + W(B),
\end{eqnarray*}
and, therefore, we have a decomposition
\[
T = S(A) + S(B) + W,
\]
where $W$ is given by
\begin{eqnarray*}
W &=T - \bigl(S(A) + S(B)\bigr) = W(A) + W(B) -T.
\end{eqnarray*}
It is easy to check that $W \geq0$, and that $W$ has the form of
a weighted Gini mean on distances between pairs of
observations in cells $\{A_i \cap B_j\}$, $1 \leq i \leq a$, $1 \leq j
\leq b$.

Similarly, we can also decompose total dispersion on factor $A\dvtx B$ to obtain
\mbox{$ T = S(A:B) + W(A:B)$}. The between component
$S({A\dvtx B})$ contains the between distances on factor $A$ and the
between distances on factor $B$. It can be shown that $S(A:B)-S(A)-S(B) \geq0$ by a similar argument as
in the proof of Corollary~\ref{c1}(i). Hence, we can obtain the decomposition
%
\begin{equation} \label{SAB}
T = S(A) + S(B) + S({AB}) + W(A:B),
\end{equation}
where $S(AB)=S(A:B)-S(A)-S(B)$.

\subsection{The DISCO decomposition for general factorial designs}

By induction, it follows that for additive models with $k \geq1$
factors and no interactions, the total dispersion can be decomposed as
\[
T = \sum_{j=1}^k S(j) + W,
\]
where $W$ is given by
%
\begin{eqnarray}\label{www}
W &=& \sum_{j=1}^k W({j}) - (k-1) T,
\end{eqnarray}
$W \geq0$, and $W$ has the form of a Gini mean on distances
between observations. [For simplicity we drop the factor label and use
a number to identify the factor in $S(j)$ and $W(j)$.]

For models with interaction terms, we proceed as in (\ref{SAB}). For a
factorial design on three factors ($A,B,C$), the highest order
interaction is $A\dvtx B\dvtx C$. In the decomposition $T=S(A\dvtx B\dvtx C)+W(A\dvtx B\dvtx C)$, the
between component $S(A\dvtx B\dvtx C)$ contains between distances for lower order
terms. Define $S(ABC)$ by
%
\begin{eqnarray}\label{SABC}
S(ABC) &=& S(A\dvtx B\dvtx C) -[S(A\dvtx B)+S(A\dvtx C)+S(B\dvtx C)] \nonumber\\
&& {}+ [S(A)+S(B)+S(C)]\\
& =&  S(A\dvtx B\dvtx C)  -  [S(AB)+S(AC)+S(BC)+S(A)+S(B)+S(C)], \nonumber
\end{eqnarray}
where $S(AB)$, $S(AC)$, and $S(BC)$ are defined as in (\ref{SAB}).
Then we obtain the decomposition shown in Table \ref{t3}.

\begin{table}
\caption{DISCO analysis for three-factor model}\label{t3}
\begin{tabular*}{\textwidth}{@{\extracolsep{\fill}}lccc@{}}
\hline
\textbf{Factor} & \textbf{df} & \textbf{Dispersion} & $\bolds{F_\alpha}$\\
\hline
A & $a-1$ & $S_A$ & $[S_A/df(A)]  /   [W / f]$ \\
B & $b-1$ & $S_B$ & $[S_B/df(B)]  /   [W / f]$ \\
C & $c-1$ & $S_C$ & $[S_C/df(C)]  /   [W / f]$ \\
AB & $(a-1)(b-1)$ & $S_{AB}$ & $[S_{AB}/df(AB)]  /   [W / f]$ \\
AC & $(a-1)(c-1)$ & $S_{AC}$ & $[S_{AC}/df(AC)]  /   [W / f]$ \\
BC & $(b-1)(c-1)$ & $S_{BC}$ & $[S_{BC}/df(BC)]  /   [W / f]$ \\
ABC & $(a-1)(b-1)(c-1)$ & $S_{ABC}$ & $[S_{ABC}/df(ABC)]  /   [W /f]$ \\
Error & $f$\tabnoteref[\dag]{tc} & $W$\tabnoteref[\ddag]{tb} & \\[5pt]
Total & $N-1$ & $T$ & \\
\hline
\end{tabular*}
\tabnotetext[\dag]{tc}{In the balanced design $f = abc(n-1)$.}
\tabnotetext[\ddag]{tb}{$W = W(A:B:C)$.}
\tabnotetext[]{ta}{$S_A=S(A)$, $S_{AB}=S(AB)$, etc.}
\end{table}

Factorial designs on four or more factors are handled in a similar way,
by obtaining $W$ from the decomposition on the highest order
interaction term, and splitting the between component into components
corresponding to the terms in the model.

Degrees of freedom are determined by the combined constraints on sums
of distances, as in linear models. The $F_\alpha$ ratios for the
$j$th term with $a_j$ levels in an additive model are
\[
F_\alpha(j) = \frac{S_\alpha(j)/(a_j-1)}{W/df(W)},
\]
where $df(W)$ equals residual degrees of freedom in the corresponding
linear model.

\section{Implementation and examples\label{s:ie}}

DISCO decomposition is easily implemented by computing the Gini sums
$G$ from the distance matrix of the sample for each of the
cells in the model. Each of the components in the decomposition is
a function of these sums.

\subsection{Calculation of test statistics} Consider the model $Y \sim
 A$ where factor $A$ has $a$ levels, corresponding to samples $A_1,A_2,\dots
,A_a$. If $\mathcal{D}$ is the $N \times N$ distance matrix of the
sample, let $M$ be the $N \times a$ design matrix defined by
\[
M = (M_{ij})= (I\{x_i \in A_j\}) = \cases{
1, & \quad $x_i \in A_j$; \cr
0, &\quad  otherwise.}
\]
Then $\mathcal G=M^T \mathcal D M$ is the $a \times a$ matrix of Gini
sums $G(A_i,A_j)$, and
the within-sample sums are along the diagonal of $\mathcal G$. Thus,
both $T$ and $W(A)$ are easily computed from $\mathcal G$, and $S(A)=T-W(A)$.

\begin{remark}
The design matrix $M$ has no intercept column and
has one column for each level of factor $A$, unlike the matrix used to fit
a linear model in most software packages. For a one-way layout the
matrix $M$ is easily obtained by software; for example, the $R$ \textit{model.matrix} function returns the required matrix $M$ for the formula
$Y \sim 0 + A$ (no intercept model).
\end{remark}

It is clear from (\ref{SAB})--(\ref{SABC}) and the example in Table
\ref{t3} that all of the required distance components for any given
model can be computed by expanding the model formula to additive form
and iteratively computing the decomposition on each term.

The calculations for a multivariate response or general $\alpha$
differ only in the initial step to compute the distance matrix
$\mathcal D$.

The DISCO test can be implemented as a permutation test, as outlined in
Section~\ref{permtest}. We have implemented DISCO tests in the
statistical computing software R (\cite{R}). The methods implemented in
this paper are available in the \textit{disco} or \textit{energy} package
for R
(\cite{disco}).

\subsection{Application: decomposition of residuals}

Suppose we consider the residuals from a fitted linear model on a
univariate response with one factor. Denote the fitted model $L$.
Regardless of whether the hypothesis of equal means is true or false,
the residuals do not reflect differences in means. If treatments differ
in some way other than the mean response, then the differences can be
measured on the residuals by distance components, $0< \alpha< 2$. If
we consider models of the type proposed by \cite{aa94}, we could
regard the linear portion $L$ for treatment effect as an ``intercept''
term. That is,
\[
F_j(x) = L(x) + R_j(x), \qquad\sum_{j=1}^a R_j(x) = 0,
\]
where $F_j$ is the distribution function of $x_{ij},  i=1,\dots,n_j$.
If all $R_j(x)=0$, then $F_j=L$ for every $j$. One can test the
hypothesis $H_0: \mathit{ all  }\ R_j(x)=0$ by testing the sample of
residuals of $L$ for equal distributions.

The following example illustrates our Theorems \ref{Th1} and \ref
{decomp}. Then DISCO decomposition is applied to the residuals.

\begin{example}[(Gravity data)]\label{ex1}
The gravity data consist of 81 measurements in a series of eight
experiments conducted by the National Bureau of Standards in Washington
DC between May, 1934 and July, 1935, to estimate the acceleration due
to gravity at Washington. Each experiment consisted of replicated
measurements with a reversible pendulum expressed as deviations from
980 cm$/$sec$^2$. The~data set (\textit{gravity}) is discussed in Example
3.2 of \citeauthor{dh97} (\citeyear{dh97}) and is available in the \textit{boot} package for R
(\cite{boot}). Boxplots of the data in Figure \ref{gravity} reveal
nonconstant variance of the measurements over the series of experiments.

\begin{figure}

\includegraphics{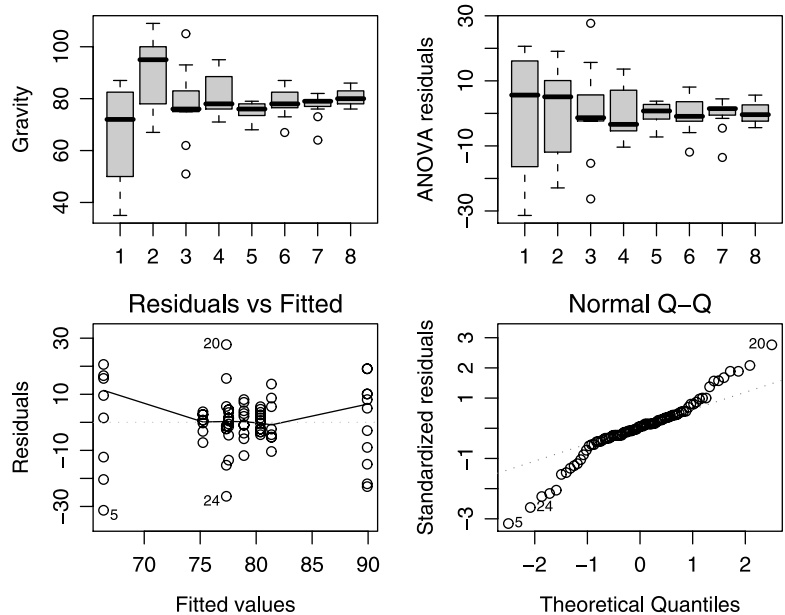}

\caption{Gravity data and residual plots in Example \protect\ref{ex1}
(sample sizes 8, 11,  9,  8,  8, 11, 13, and 13).}\label{gravity}
\end{figure}

The decompositions by series for $\alpha=1$ and $\alpha=2$ are shown
in Table \ref{tab1}. Note that when index $\alpha=2$ is applied, the
DISCO decomposition is exactly equal to the ANOVA decomposition, also
shown in Table \ref{tab1}. In fact, with our implementation as random
permutation test, the $F_2$ test is actually a \textit{permutation test}
based on the ANOVA $F$ statistic. In this example 999 permutation
replicates were used to estimate the $p$-values.

\begin{table}
\caption{Comparison of DISCO and ANOVA decompositions in Example 1}\label{tab1}
\rule{\textwidth}{0.6pt}
\begin{verbatim}
DISCO
  Distance Components: index 1.00
  Source            Df   Sum Dist  Mean Dist    F-ratio    p-value
  Between:
    Series           7  100.62287   14.37470      2.781      0.001
  Within            73  377.27836    5.16820

  Distance Components: index 2.00
  Source            Df   Sum Dist  Mean Dist    F-ratio    p-value
  Between:
    Series           7  2818.62413 402.66059      3.568      0.002
  Within            73  8239.37587 112.86816

ANOVA
  Analysis of Variance Table
  Response: Gravity
            Df Sum Sq Mean Sq F value  Pr(>F)
  Series     7 2818.6  402.7  3.5675 0.002357 [0.002 by perm. test]
  Residuals 73 8239.4  112.9
\end{verbatim}
\rule{\textwidth}{0.6pt}
\end{table}

Residual plots from the fitted linear model (ANOVA) are shown in Figure
\ref{gravity}, indicating that residuals have non-normal distribution
and nonconstant variance. When we decompose residuals by Series using
DISCO ($\alpha=1$) as shown in Table~\ref{tab1a}, the DISCO $F_1$
statistic is significant ($p$-value $< 0.05$). We can conclude that the
residuals do not arise from a common error distribution. (The ANOVA $F$
statistic is zero on residuals.)
\end{example}

\begin{table}[b]
\caption{Distance Components of ANOVA residuals in Example 1}\label{tab1a}
\rule{\textwidth}{0.6pt}
\begin{verbatim}
Distance Components: index  1.00
Source            Df   Sum Dist  Mean Dist   F-ratio   p-value
Between:
  Series           7   56.66334    8.09476     1.566     0.046
Within            73  377.27836    5.16820
\end{verbatim}
\rule{\textwidth}{0.6pt}
\end{table}

The next example illustrates decomposition of residuals for a
multivariate response.

\begin{example}[(Iris data)]\label{iris}
Fisher's (or Anderson's) iris data set records four measurements
(sepal length and width, petal length and width) for 50 flowers
from each of three species of iris. The species are iris setosa,
versicolor, and virginica. The data set is available in R (\textit{iris}).
The model is $Y \sim$ Species, where $Y$ is a four dimensional response
corresponding to the four measurements of each iris. The
DISCO $F_1$ and MANOVA Pillai--Bartlett $F$ test, implemented as
permutation tests, each have $p$-value 0.001 based on 999 permutation
replicates. The residuals from the fitted linear model are a $150
\times4$ data set.

Results of the multivariate analysis are shown in Table \ref{tabiris}.
From the DISCO decomposition of the residuals and test for equality of
distributions of residuals ($p$-value $< 0.04$), it appears that there
are differences due to Species that are not explained by the linear
component of the model.
\end{example}

\begin{table}
\caption{Analysis of iris data and residuals in Example \protect\ref{iris}}\label{tabiris}
\rule{\textwidth}{0.6pt}
\begin{verbatim}
DISCO analysis of multivariate iris data:
    Distance Components: index   1.00
    Source           Df   Sum Dist   Mean Dist   F-ratio   p-value
    Between:
      Species         2  119.23731    59.61865   124.597     0.001
    Within          147   70.33848     0.47849

MANOVA analysis of multivariate iris data:
               Df Pillai approx F num Df den Df    Pr(>F)
   Species      2  1.192   53.466      8    290 < 2.2e-16 ***
   Residuals  147
   [permutation test p = 0.001]

DISCO analysis of residuals of linear model for iris data:
    Distance Components: index   1.00
    Source           Df   Sum Dist   Mean Dist   F-ratio   p-value
    Between:
      Species         2    1.69845     0.84923     1.775     0.039
    Within          147   70.33848     0.47849
\end{verbatim}
\rule{\textwidth}{0.6pt}
\end{table}
%

\subsection{Choosing the index $\alpha$}

Choice of a test or a parameter for a test is a difficult question.
Consider the similar situation one has with the choice of Cram\'
{e}r--von Mises tests, an infinite class of statistics that depend on
the choice of weight function. For testing normality, for example, one
can use the identity weight function (Cram\'{e}r--von Mises test) or
weight function $F(x)(1-F(x))$ (Anderson--Darling test) and both are
good tests with somewhat different properties. Here we have a similar choice.

The simplest and most natural choice is $\alpha=1$ corresponding to
Euclidean distance. It is natural because it is at the center of our
interval for $\alpha$.
Considering implementation for a univariate response, when $\alpha=1$ the
Gini means can be linearized, which reduces the computational
complexity from $O(N^2)$ to $O(N \log(N))$.

For heavy-tailed distributions one may want to apply a small $\alpha$,
which could be selected based on the data. As an example, consider the
Pareto distribution with density $f(x) = k \sigma^k / x^{k+1}$, $x >
\sigma$. In this case $E[X]$ exists only for $k > 1$ and $\operatorname{Var}(X)$ is finite only for $k > 2$. Note that $X^\alpha$ has a
Pareto distribution for $\alpha>0$. If one is comparing claims data,
which Pareto models tend to fit well, the tail index~$k$ can be
estimated by maximum likelihood to find a conservative choice of
$\alpha$ such that the second moments of $X^\alpha$ exist.
Heavy-tailed stable distributions such as L\'{e}vy distributions used
in financial modeling suggest another situation where $\alpha<1$ may
be recommended.

\section{Simulation results}\label{empirical}

In this section we present the results of Monte Carlo
studies to assess power of DISCO tests. In our simulations $R=199$
replicates are generated for each DISCO test decision.

Examples \ref{ex5} and
\ref{ex6} compare DISCO with two parametric MANOVA tests based on
\cite{pillai55} and \cite{wilks32} statistics (see, e.g.,
\citeauthor{anderson84} (\citeyear{anderson84}, Chapter~8)). The Pillai--Bartlett test implemented
in R is recommended by \cite{ht87}.

\begin{example}\label{ex5}
The multivariate response is generated in a
four group balanced
design with common sample size $n=30$. The marginal distributions
are independent with Student $t(4)$ distributions. Sample 1 is
noncentral $t(4)$ with noncentrality parameter $\delta$. Samples
2--4 each have central $t(4)$ distributions. The index applied in the
DISCO test is 1.0.

Results of several simulations are summarized in Figure
\ref{fig2}(a) and (b) at significance level $0.10$. In Figure
\ref{fig2}(a) the noncentrality parameter is on the horizontal axis
and dimension is fixed at $p=10$. In Figure \ref{fig2}(b) the
dimension is on the horizontal axis and $\delta=0.2$ is fixed. Each
test achieves approximately the nominal significance level of $10\%$
under the null hypothesis [see Figure \ref{fig2}(a) at $\delta=0$].
Standard error of the estimate of power is at most 0.005, based on
10,000 tests.

\begin{figure}
\tabcolsep=4pt
\begin{tabular*}{\textwidth}{@{\extracolsep{\fill}}cc@{}}

\includegraphics{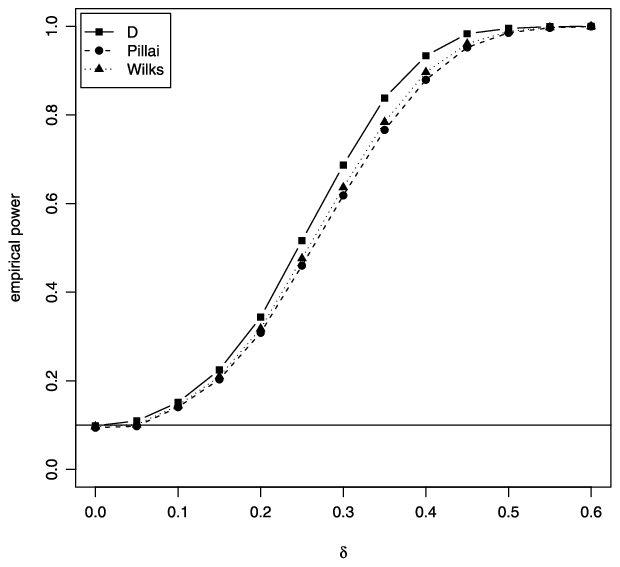}
&\includegraphics{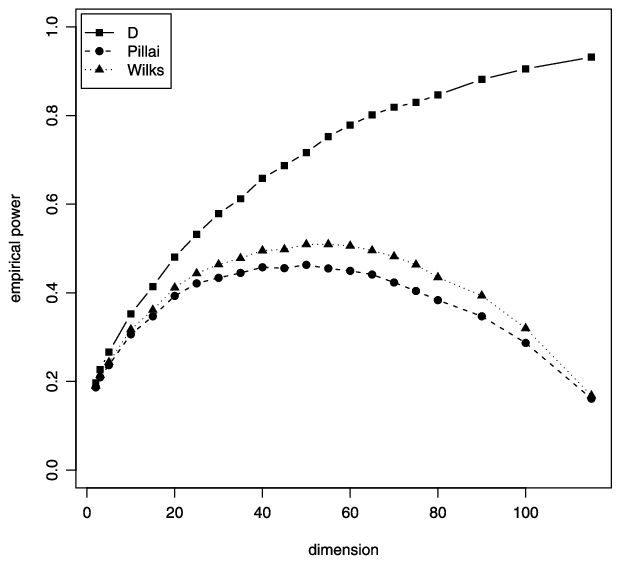}\\
\footnotesize{(a)}&\footnotesize{(b)}
\end{tabular*}
\caption{Monte Carlo results for Example \protect\ref{ex5}:
 Empirical power of the DISCO and MANOVA tests against
 a t(4) alternative, four groups with $n=30$ per group,
 where \textup{(a)} dimension $p=10$ and noncentrality
 parameter $\delta$ varies and \textup{(b)}
 $p$ varies and $\delta=0.2$. Standard error of power estimate
 is at most 0.005.}\label{fig2}
\end{figure}

Results displayed in Figure \ref{fig2}(a) and (b) suggest that the DISCO test
is slightly more powerful than MANOVA tests against this alternative
when $p=10$. As dimension increases, Figure \ref{fig2}(b) illustrates
that the DISCO test is increasingly superior relative to MANOVA
tests.

The MANOVA tests apply a transformation to obtain an approximate $F$
statistic. Although the data is non-normal, the MANOVA test
statistics appear to be robust to non-normality in this example and
exhibit good power when $p=10$. This simulation suggests that the
transformation may not be applicable for test decisions when
dimension is large relative to number of observations. For
comparison with MANOVA tests, dimension is constrained by sample
size. Note, however, that the DISCO test is applicable in arbitrary
dimension regardless of sample size.
\end{example}

\begin{example}\label{ex6}
In this example we again consider a balanced
design with four groups
and $n=30$ observations per group. Groups 2--4 have i.i.d. marginal
Gamma(shape${}={}$2, rate${}={}$0.1) distributions. Group 1 is
also\break
Gamma(shape${}={}$2, rate${}={}$0.1), but with multiplicative errors distributed
as Lognormal($\mu=0, \sigma$). Thus, the natural logarithm of the
group 1 response has an additive normally distributed error with
mean 0 and variance $\sigma^2$. The index applied is 1.0.

Results for significance level $10\%$ are summarized in Figures
\ref{fig3}(a) and (b). Each test achieves approximately the
nominal significance level of $10\%$ under the null hypothesis [see
Figure \ref{fig3}(a) at $\sigma=0$]. Standard error of the estimate
of power is at most 0.005, based on 10,000 tests.

\begin{figure}
\tabcolsep=4pt
\begin{tabular*}{\textwidth}{@{\extracolsep{\fill}}cc@{}}

\includegraphics{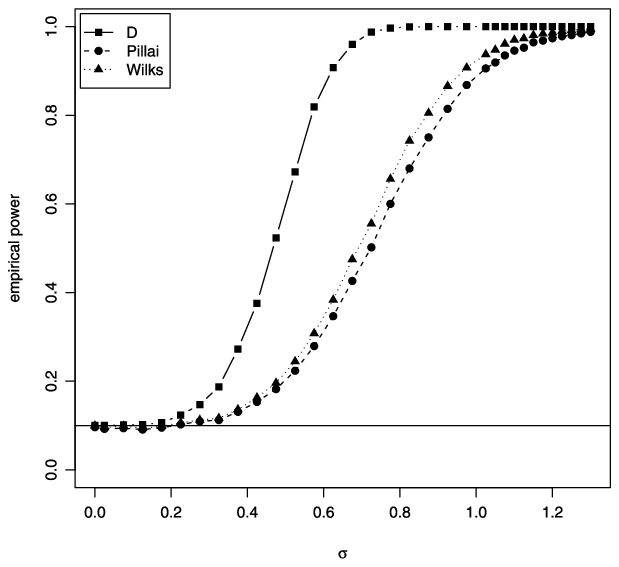}
&\includegraphics{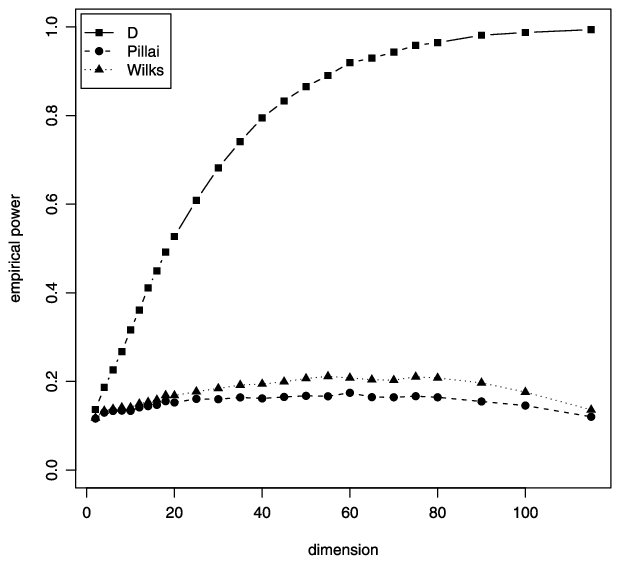}\\
\footnotesize{(a)}&\footnotesize{(b)}
\end{tabular*}
\caption{Monte Carlo results for Example \protect\ref{ex6}:
Empirical power of the DISCO and MANOVA tests against
a gamma(shape${}={}$2, rate${}={}$0.1) alternative, four groups with $n=30$ per group,
where \textup{(a)}~dimension $p=10$ and $\sigma$ varies \textup{(b)}
$p$ varies and $\sigma=0.4$. Standard error of power estimate is at
most~0.005.}\label{fig3}
\end{figure}

In Figure \ref{fig3}(a) the parameter $\sigma$ is on the horizontal
axis and dimension is fixed at $p=10$. Each test exhibits empirical
power increasing with $\sigma$ in Figure \ref{fig3}(a), but the DISCO
test is clearly more powerful than the MANOVA tests against this
alternative. In Figure \ref{fig3}(b) the dimension is on the
horizontal axis and $\sigma=0.4$ is fixed. This simulation reveals
increasingly superior power performance of DISCO as dimension
increases.
\end{example}

\section{Summary}\label{summary}

The distance components decomposition of total dispersion is
analogous to the classical decomposition of variance, but
generalizes the decomposition to a family of methods indexed by an
exponent in $(0, 2]$. The ANOVA and MANOVA methods are extended by
choosing an index strictly less than 2, for which we obtain a
statistically consistent test of the general hypothesis of equal
distributions. DISCO tests can be applied in arbitrary dimension,
which is not constrained by number of observations. The usual
assumption of homogeneity of error variance is not required for
DISCO tests, and the distribution of errors need not be specified
except for the mild condition of finite variance. Moreover, the
DISCO permutation test implementation is nonparametric and does not
depend on the distributions of the sampled populations.

\appendix

\section{}\label{append}

\subsection{\texorpdfstring{Proof of Proposition
\protect\ref{P:sst2}}{Proof of Proposition 1}}

The total sum of squared distances can be decomposed as
\begin{eqnarray*}
\sum\limits_{m=1}^{n_2}
\sum\limits_{i=1}^{n_1} |a_i-b_m|^2 &=&
\sum\limits_{m=1}^{n_2}
\sum\limits_{i=1}^{n_1} |a_i-\bar a + \bar a - b_m|^2\\
&=& \sum\limits_{m=1}^{n_2} [
n_1 \hat\sigma_1^2 + n_1(\bar a - b_m)^2] \\ &=&
n_1 n_2 \hat\sigma_1^2 + n_1 \sum\limits_{m=1}^{n_2}
(\bar a -\bar b + \bar b - b_m)^2 \\&
=& n_1 n_2 [\hat\sigma_1^2 + \hat\sigma_2^2 + (\bar a - \bar
b)^2],
\end{eqnarray*}
where $\hat\sigma_1^2=(1/n_1)\sum_{i=1}^{n_1}(a_i-\bar a)^2$ and
$\hat\sigma_2^2=(1/n_2)\sum_{i=1}^{n_2}(b_i-\bar b)^2$. Similarly,
\begin{eqnarray*}
\sum\limits_{m=1}^{n_1}
\sum\limits_{i=1}^{n_1} |a_i-a_m|^2 = 2 n_1^2 \hat\sigma_1^2
\quad\mbox{and} \quad
\sum\limits_{m=1}^{n_2}
\sum\limits_{i=1}^{n_2} |b_i-b_m|^2 = 2 n_2^2 \hat\sigma_2^2,
\end{eqnarray*}
so that
\begin{eqnarray}\label{id:d2}
d_2(A,B) &=& \frac{n_1n_2}{n_1+n_2} \bigg[\frac{2}{n_1n_2}  n_1 n_2
\bigl(\hat\sigma_1^2 + \hat\sigma_2^2 + (\bar a - \bar b)^2\bigr) -
\frac{1}{n_1^2} n_1^2 \hat\sigma_1^2 -
\frac{1}{n_2^2} n_2^2 \hat\sigma_2^2 \bigg]\nonumber \\[-8pt]\\[-8pt]
&=& \frac{2n_1n_2}{n_1+n_2} (\bar a - \bar b)^2.\nonumber
\end{eqnarray}
The well-known identity
%
\begin{equation}
n_1(\bar a - \bar c)^2 + n_2(\bar b - \bar c)^2 =
\frac{n_1n_2}{n_1+n_2} (\bar a - \bar b)^2
\label{sst2}
\end{equation}
follows from $\bar a - \bar c = n_2(\bar a - \bar b)/(n_1+n_2)$ and
$ \bar b - \bar c = n_1(\bar b - \bar a)/(n_1+n_2)$. Hence, $
d_2(A,B) = 2n_1(\bar a - \bar c)^2 + 2n_2(\bar b - \bar c)^2 =
2 \mathit{SST} $.

\subsection{\texorpdfstring{Proof of Theorem \protect\ref{decomp}}{Proof of Theorem 2}}

One can obtain the DISCO decomposition by directly computing the
difference between the total and within-sample dispersion. Given
$p$-dimensional samples $A_1,\dots,A_K$ with respective sample sizes
$n_1,\dots,n_K$ and $N=\sum_j n_j$, let $g_{jk}=g_\alpha(A_j,A_k)$
given by (\ref{g}) and $G_{jk}=n_j n_k g_{jk}$, for $j,k=1,\dots,K$.
Then for all $0 < \alpha\leq2$ and $p \geq1$,
\begin{eqnarray*}
T_\alpha- W_\alpha
&=& \frac N2 g(A,A) - \frac12 \sum_j n_j g_{jj} \\
&=&
\frac N2 \biggl( \sum_{j<k} \frac2{N^2} G_{jk}
+ \sum_j \frac1{N^2} G_{jj} \biggr) - \frac12 \sum_j \frac1{n_j} G_{jj}
\\&=&
\frac1{2N} \biggl( \sum_{j<k} 2 G_{jk}
+ \sum_j G_{jj} \biggr) - \frac12 \sum_j \frac1{n_j} G_{jj}
\\&=&
\frac{1}{2N} \biggl( \sum_{j<k} n_jn_k (2g_{jk} - g_{jj} - g_{kk}
) + \sum_{j<k} n_j n_k (g_{jj}+g_{kk}) \biggr) \\
&& {}+
\frac{1}{2N} \sum_j n_j^2 g_{jj} - \frac12 \sum_j n_j g_{jj}
\\&=&
\sum_{j<k} \frac{n_j+n_k}{2N} \biggl(\frac{n_j n_k}{n_j + n_k}
\biggr) (2 g_{jk} - g_{jj} - g_{kk} ) + \frac{1}{2N} \sum_{j<k}
n_k(n_j g_{jj}) \\
&&
{}+ \frac{1}{2N} \sum_{j<k} n_j(n_k g_{kk})
+\frac{1}{2N} \sum_j n_j^2 g_{jj} - \frac12 \sum_j n_j g_{jj}.
\end{eqnarray*}
After simplification we have
\begin{eqnarray*}
T_\alpha- W_\alpha&=&
\sum_{j<k} \frac{n_j+n_k}{2N} \biggl(\frac{n_j n_k}{n_j + n_k}
\biggr) (2 g_{jk} - g_{jj} - g_{kk} ) \\&
&{}+ \frac{1}{2N} \sum_{k} \sum_{j} n_k (n_j g_{jj})
- \frac12 \sum_j n_j g_{jj} \\&=&
\sum_{j<k} \frac{n_j+n_k}{2N} \biggl(\frac{n_j n_k}{n_j + n_k}
\biggr) (2 g_{jk} - g_{jj} - g_{kk} ) \\
&&{}+ \frac{N}{2N} \sum_{j} n_j g_{jj}
- \frac12 \sum_j n_j g_{jj} \\&=&
\sum_{j<k} \frac{n_j+n_k}{2N} \biggl(\frac{n_j n_k}{n_j + n_k}
\biggr) (2 g_{jk} - g_{jj} - g_{kk} ) = B_\alpha.
\end{eqnarray*}

\printaddresses

\end{document}